\documentclass[12pt,showpacs,superscriptaddress]{revtex4}
\usepackage{amsmath,amssymb,amsthm,amsfonts}
\usepackage{hyperref}

\begin{document}

\title{Hybrid Quantum Cosmology: Combining Loop and Fock Quantizations}

\author{G. A. Mena Marug\'{a}n}\email{mena@iem.cfmac.csic.es}
\affiliation{Instituto de Estructura de la Materia, CSIC, Serrano
121, 28006 Madrid, Spain}

\author{M. Mart\'{i}n-Benito}
\email{merce.martin@iem.cfmac.csic.es} \affiliation{Instituto de
Estructura de la Materia, CSIC, Serrano 121, 28006 Madrid, Spain}

\begin{abstract}

As a necessary step towards the extraction of
realistic results from Loop Quantum Cosmology, we
analyze the physical consequences of including
inhomogeneities. We consider in detail the
quantization of a gravitational model in vacuo which
possesses local degrees of freedom, namely, the
linearly polarized Gowdy cosmologies with the spatial
topology of a three-torus. We carry out a hybrid
quantization which combines loop and Fock techniques.
We discuss the main aspects and results of this hybrid
quantization, which include the resolution of the
cosmological singularity, the polymeric quantization
of the internal time, a rigorous definition of the
quantum constraints and the construction of their
solutions, the Hilbert structure of the physical
states, and the recovery of a conventional Fock
quantization for the inhomogeneities.

\end{abstract}

\pacs{04.60.Pp, 98.80.Qc, 04.62.+v}

\maketitle
\newpage

\section{Introduction}
\setcounter{subsection}{0}

In spite of the impressive progress that Cosmology has
experienced in recent years, we are still missing a
consistent explanation of the origin of the Universe
and the formation of structures which is deduced
entirely from a fundamental theory. General Relativity
(GR) breaks down in the very initial instants of the
history of the Universe, leading to a cosmological
singularity of the big bang type \cite{hawel}. In this
regime GR cannot be trusted, and the very own
predictability of the laws of physics is lost. One
expects instead that the physics of the Primitive
Universe belongs to the realm of Quantum Gravity,
namely, a theory of the gravitational field which
incorporates the quantum behavior of nature. One of
the most promising candidates for such a theory is
Loop Quantum Gravity \cite{lqg}. At present, important
efforts are being made in order to adapt the
techniques of Loop Quantum Gravity to much simpler
settings than those of the complete theory, which on
the other hand remains to be concluded. This is the
case of a series of cosmological models obtained from
GR by symmetry reduction. The resulting field of
research is known under the general name of Loop
Quantum Cosmology (LQC) \cite{lqc}.

The first cosmological system whose quantization was
performed to completion in LQC was the homogeneous,
isotropic, and spatially flat model provided with a
minimally coupled, homogeneous, and massless scalar
field as matter content \cite{abl,aps1,aps3}.
The geometry was polymerically quantized (i.e., using
typical LQC methods), while the matter field was
described using standard quantization methods. A
thorough analysis of the resulting quantum
dynamics \cite{aps1,aps3} showed that the initial
singularity is successfully resolved. The classical
big bang is replaced with a quantum big bounce which
deterministically connects a semiclassical expanding
universe with a previous semiclassical contracting
one. After the pioneer study of this simple model, LQC
has been further developed to describe other
homogeneous isotropic systems \cite{iso}, or homogenous
and anisotropic cosmologies like, e.g., the Bianchi I
model \cite{chio,mmp,szu}. In all these works,
the cosmological singularity is eluded in the quantum
dynamics, a result which strengthens the validity of
the singularity resolution mechanism and the relevance
of LQC.

One may wonder whether the quantum resolution of the
singularities of GR is just an artifact of the great
symmetry of the models analyzed so far in the
literature, and in particular of the homogeneity, or
whether, on the contrary, singularities are removed as
well when inhomogeneities are present. In order to ask
this question, it seems unavoidable to extend the
quantization performed in LQC to inhomogeneous
scenarios. Moreover, since inhomogeneities may have
played a crucial role at the first instants of the
Universe, one should take them into account, indeed,
in the development of a realistic theory of quantum
cosmology.

In order to progress in this direction, we will
consider here the quantization of one of the simplest
inhomogeneous cosmological systems, namely the
linearly polarized Gowdy $T^3$ model \cite{gowd}. This
model is a natural test bed to incorporate
inhomogeneities in LQC. On the one hand, its
quantization by means of standard techniques has been
discussed in detail \cite{qGow}, and a successful Fock
(and Schr\"odinger) quantization has already been
achieved \cite{men1,men2}. Even though this system has
no timelike isometry and possesses an infinite number
of degrees of freedom, this quantization has been
shown to be unique (up to unitary equivalence) under
certain reasonable requirements, namely, it is the
unique Fock quantization with a unitary implementation
of the dynamics and invariance under the action of
diffeomorphisms (strictly speaking, of the only
diffeomorphisms which remain in the system after a
natural gauge fixing: the $S^1$ translations). On the
other hand, generically, the classical solutions of
the Gowdy model represent spacetimes with an initial
curvature singularity \cite{mon,ise}. Besides, the
subset formed by the homogeneous solutions represents
Bianchi I spacetimes with a three-torus topology, a
model which has been polymerically quantized and where
the singularity has again been shown to be resolved as
a result of the quantization \cite{mmp,szu}. Therefore,
it is natural to ask how the inclusion of the
inhomogeneities affects the singularity resolution in
this particular model. In view of the knowledge
available about the system and the characteristics of
the different gravitational degrees of freedom
entering the model, the simplest possibility to
investigate that question is to perform a hybrid
quantization, which combines the polymeric
quantization of the homogeneous degrees of freedom
with the Fock quantization of the inhomogeneities.
This idea constitutes the basis of our work, whose
main aspects have already been presented in Ref.
\cite{let}. Actually, this hybrid quantization will
allow us to investigate not only whether the
singularity is resolved owing to quantum geometry
effects, but, furthermore, to explore whether the loop
quantization of just the degrees of freedom that
parametrize the homogeneous solutions (zero modes in
our description) may suffice to cure the big bang
singularity. As we will see, the answer turns out to
be in the affirmative for our model. The simplicity of
this result reinforces the interest of our hybrid
approach, since the same strategy can be applied to
the quantization of more general cosmologies.

In addition, the hybrid quantization of the Gowdy
model provides a specially suitable arena for the
analysis of other important issues in quantum gravity
and cosmology, such as:
\begin{itemize}
\item The role of the internal time.
We will select as emergent time a gravitational
variable which behaves like a clock in the classical
theory. However, in contrast with the situation
studied in the recent literature \cite{aps1,aps3}, we
will quantize this emergent time using a polymeric
representation, because it describes one of the
degrees of freedom of the homogeneous sector of
solutions. As a consequence of this polymeric
quantization, we will see that, in principle, the
evolution with respect to that time is not unitary.
\item The recovery of the standard quantum field theory.
One can regard the inhomogeneities as (conveniently
scaled) gravitational waves propagating in a
homogeneous curved background, namely a Bianchi I
spacetime. In a standard quantization, the
inhomogeneities can be described by means of a Fock
space. One may then ask what is the status of this
Fock description from the viewpoint of LQC, where one
adopts a polymeric quantization inequivalent to the
standard one. We will see that, with our hybrid
approach, one indeed recovers the standard Fock
description of the inhomogeneities, even though the
background is quantized employing loop techniques.
\end{itemize}
The rest of the paper is organized as follows. In Sec.
2 we will introduce the classical system. We will
partially fix the gauge and parametrize the degrees of
freedom of the homogeneous solutions (what we will
call the homogeneous sector) in Ashtekar variables,
whereas the inhomogeneities will be described by
nonzero Fourier modes of a certain scalar field (this
will be called the inhomogenous sector). A global
diffeomorphism constraint remains in the reduced
model, affecting only the inhomogeneous sector. This
sector will be quantized in Sec. 3 adopting a Fock
representation. In addition, we will impose the
quantum analog of the mentioned diffeomorphism
constraint. On the other hand, a global Hamiltonian
constraint, coupling the homogeneous and inhomogeneous
sectors, is also present. In Sec. 4 we will construct
a polymeric representation of the homogeneous sector,
which coincides with the phase space of a vacuum
Bianchi I model. The Hamiltonian constraint of the
complete system will be imposed in Sec. 5, where we
will also construct the physical Hilbert space of the
hybrid Gowdy model. Finally, in Sec. 6 we will discuss
the main results of our quantization.

\section{The Classical Gowdy Model}
\setcounter{subsection}{0}

The Gowdy cosmologies are vacuum spacetimes with two
spacelike commuting Killing vector fields and spatial
sections of compact topology \cite{gowd}. We consider
the simplest of these Gowdy models, namely, the case
with the spatial topology of a three-torus, $T^3$, and
with a linear polarization. In this case both Killing
vectors are axial (because of the topology) and
hypersurface orthogonal (owing to the considered
polarization).

Let us denote by $\partial_\sigma$ and
$\partial_\delta$ the two Killing vector fields. To
describe the system, we choose global coordinates
$\{t,\theta,\sigma,\delta\}$ adapted to the
symmetries, with $\theta,\sigma,\delta\in S^1$. The
metric components depend on $t$ and $\theta$, being
periodic in the latter coordinate. The spacetime is
globally hyperbolic and, in a 3+1 decomposition, the
metric can be described by the densitized lapse
function ${N_{_{_{\!\!\!\!\!\!\sim}}\;}}$, the shift
vector $N^{i}$, and the three-dimensional metric
$q_{ij}$ induced on the spatial slices which foliate
the manifold, where $i,j\in\{\theta, \sigma,\delta\}$.
As a consequence of the hypersurface orthogonality,
the induced metric is such that $q_{\sigma\delta}=0$.
Besides, the conditions that $q_{\theta\sigma}$ and
$q_{\theta\delta}$ vanish fix the gauge freedom
associated with the diffeomorphism constraints in
$\sigma$ and $\delta$, the directions defined by the
Killing vector fields \cite{man}. In addition, the
dynamical stability of these conditions imply that
$N^\sigma$ and $N^\delta$ must be equal to zero. At
this stage, the induced metric is diagonal and
therefore can be characterized by three fields that
describe the norm of one of the Killing vectors, the
area of the isometry group orbits, and the scale
factor of the metric induced on the set of group
orbits. We further demand that both the generator of
the conformal transformations of this latter metric
and the area of the isometry group orbits be
homogeneous functions. These conditions turn out to
fix the gauge freedom associated with the nonzero
(inhomogeneous) Fourier modes (with respect to the
$\theta$-dependence) of both the diffeomorphism
constraint in the $\theta$-direction and the
densitized Hamiltonian constraint. Besides, they imply
that $N^\theta$ and ${N_{_{_{\!\!\!\!\!\!\sim}}\;}}$
must be homogeneous functions \cite{men1,new}. The
difference between this gauge fixing and reduction
procedure and the one considered in Ref.
\cite{men1} is that, in that reference, the system
was totally deparametrized, since the area of the
isometry group orbits (which is a global time
function) was entirely fixed, whereas now we leave
unfixed its homogeneous part (the zero Fourier mode).

As a result of the almost complete gauge fixing that
we have performed, all the gauge degrees of freedom
are fixed except two. The resulting reduced phase
space splits into two sectors, the homogeneous sector
and the inhomogeneous one. The former of these sectors
is formed by the degrees of freedom that describe
homogeneous metric functions in our gauge fixing,
together with their conjugate variables. This includes
the zero mode of the only metric field which has not
been fixed yet, and which provides the norm of one of
the Killing vector fields (e.g., $\partial_\delta$).
The inhomogeneous sector, on the other hand, contains
the information about all the nonzero modes of this
metric field and their conjugate momenta.

We next describe the homogeneous sector in terms of
Ashtekar variables: an $SU(2)$ gravitational
connection $A_i^a$ and a densitized triad $E_a^i$,
both of them constant on spatial sections. It is worth
commenting that this homogeneous sector can be
interpreted as the phase space of a vacuum Bianchi I
model whose compact sections are homeomorphic to a
three-torus. These are precisely the spacetimes
described by the considered degrees of freedom when
the inhomogeneities vanish. Hence, the system is
provided with a natural coordinate cell, namely the
$T^3$-cell, with sides of coordinate length equal to
$2\pi$. In a diagonal [$SU(2)$] gauge, and choosing
from now on the Euclidean metric as fiducial metric,
the nontrivial components of the corresponding
Ashtekar variables \cite{mmp,chi2} are \cite{note}
\begin{equation}
A_i^a=\frac{c^{i}}{2\pi}\delta_i^a, \qquad E_a^i=
\frac{p_{i}}{4\pi^2}\delta_a^i.
\end{equation}
The nonvanishing Poisson brackets between these
variables are $\{c^i,p_j\}=8\pi G\gamma\delta^i_j$.
Here, $a$ is an internal $SU(2)$ index, the symbol $G$
stands for the Newton constant, and $\gamma$ is the
Immirzi parameter \cite{gior}.

The rest of degrees of freedoms of our reduced system
correspond to a metric field $\xi(\theta)$ and its
conjugate momentum $P_\xi(\theta)$ \cite{men1}, both of
them devoid of any zero mode contribution since these
have already been included in the homogenous sector.
Appropriate variables for the description of what we
call the inhomogeneous sector are then all the nonzero
Fourier modes $\{(\xi_m,P_{\xi}^m),m\in\mathbb{Z}
-\{0\}\}$. Now, we introduce the creation and
annihilation variables $\{(a_m,a_m^*)\}$ which would
be naturally associated with $\xi(\theta)$ if this
were a free massless scalar field. These are defined
by
\begin{align}\label{as2}
a_m&=\sqrt{\frac{\pi}{8G|m|}}\left(|m|\xi_{_{m}}
+i\frac{4G}{\pi}P_{\xi}^{m}\right),
\end{align}
and the complex conjugate relation, and are such that
$\{a_m,a_{\tilde m}^*\}=-i\delta_{m\tilde m}$.

The explicit relation between the chosen variables and
the metric is \cite{men1,new}
\begin{equation}
ds^2=\frac{|p_\theta p_\sigma p_\delta|}{4\pi^2}
\left[e^{\tilde\gamma}\left(-\frac{
{N_{_{_{\!\!\!\!\!\!\sim}}\;}}^2}{(2\pi)^4}dt^2+
\frac{d\theta^2}{p_\theta^2}\right)
+e^{-\frac{2\pi\tilde\xi}{\sqrt{|p_\theta|}}}
\frac{d\sigma^2}{p_\sigma^2}+
e^{\frac{2\pi\tilde\xi}{\sqrt{|p_\theta|}}}
\frac{d\delta^2}{p_\delta^2}\right],
\end{equation}
where
\begin{equation}
\tilde\xi(\theta)=\sum_{m\neq0} \frac{\sqrt{G}}{\pi
\sqrt{|m|}}(a_m+a_{-m}^*)e^{im\theta},
\end{equation}
and
\begin{equation}
\tilde\gamma(\theta)= \bigg(\frac{2c_\delta
p_\delta}{c_\sigma p_\sigma+c_\delta
p_\delta}-1\bigg)\frac{2\pi}
{\sqrt{|p_\theta|}}\tilde\xi(\theta)-\frac{\pi^2}{|p_\theta|}
[\tilde\xi(\theta)]^2-\frac{8\pi G\gamma }{c_\sigma
p_\sigma+c_\delta p_\delta}\zeta(\theta),
\end{equation}
with
\begin{equation}
\zeta(\theta)=i\sum_{\tilde m,m\neq0}
\text{sign}(m+\tilde m)\frac{\sqrt{|m+\tilde m||\tilde
m|}}{m}\big(a_{-\tilde {m}} -a^*_{\tilde
m}\big)\big(a_{m+\tilde m} +a^*_{-(m+\tilde
m)}\big)e^{im\theta}.
\end{equation}
If the inhomogeneities vanish, we recover the Bianchi
I spacetime. Let us comment that $\tilde\gamma$
contains a zero mode contribution arising from
quadratic terms in the inhomogeneities. We note also
that $p_\theta$ is proportional to the time variable
chosen in Ref. \cite{men1} to deparametrize the
system.

The gauge fixing that we have performed is partial.
This explains the appearance of a homogenous
densitized lapse ${N_{_{_{\!\!\!\!\!\!\sim}}\;}}$ in
the metric, and the fact that is still allowed to
redefine $\theta$ by introducing a homogeneous shift
$N^\theta$. As a consequence, two global constraints
remain in the model. One of them is the zero Fourier
mode of the diffeomorphism constraint in $\theta$,
$C_\theta$, which generates translations in the
circle. The other is the zero mode of the densitized
Hamiltonian constraint, $\tilde C_{\text{G}}$. The
volume of the Bianchi I counterpart of the Gowdy
spacetime (for vanishing inhomogeneities), which is
given by $V=\sqrt{|p_\theta p_\sigma p_\delta|}$ and
that we will call the ``homogeneous-volume'', allows
us to define a global (nondensitized) Hamiltonian
constraint, $C_{\text{G}}$, instead of $\tilde
C_{\text{G}}$. It is obtained by choosing the
(nondensitized) homogeneous lapse function $N=
V{N_{_{_{\!\!\!\!\!\!\sim}}\;}}$ rather than
${N_{_{_{\!\!\!\!\!\!\sim}}\;}}$. The two remaining
constraints can then be expressed as \cite{let}
\begin{equation}\label{classicCtheta}
C_\theta=\sum_{m=1}^\infty m(a_m^*a_m-a_{-m}^*a_{-m})=0,
\end{equation}
\begin{equation}\label{classicCG}
C_{\text{G}}=\frac{\tilde C_{\text{G}}}{V}=
C_{\text{BI}}+C_{\xi}=0,
\end{equation}
\begin{equation}\label{classicCBI}
C_{\text{BI}}=-\frac{2}{\gamma^2V}[c_\theta p_\theta
c_\sigma p_\sigma+c_\theta p_\theta c_\delta p_\delta
+c_\sigma p_\sigma c_\delta p_\delta],
\end{equation}
\begin{equation}\label{classicCXI}
C_{\xi}=\frac{G}{V}\bigg[\frac{(c_\sigma p_\sigma+
c_\delta p_\delta)^2}{\gamma^2|p_\theta|}
H_\text{int}^\xi+32\pi^2|p_\theta|H_0^\xi\bigg].
\end{equation}
$C_{\text{BI}}$ is the classical Hamiltonian
constraint of the Bianchi I model, whereas $C_{\xi}$
is the inhomogeneous term, which couples in a
nontrivial manner the homogeneous sector of the Gowdy
model with the inhomogeneities. These are encoded in
the contributions
\begin{equation}\label{Hamiltonians}
H_0^\xi=\sum_{m\neq 0}|m|a^*_ma_m,
\qquad H_\text{int}^\xi=\sum_{m\neq 0}
\frac{1}{2|m|}\left[2a^*_ma_m+
a_ma_{-m}+a^*_ma^*_{-m}\right].
\end{equation}
These terms represent, respectively, the Hamiltonian
corresponding to a free massless scalar field and an
interaction term quadratic in the field.

The choice of variables that we have made is the most
suitable one for the subsequent hybrid quantization
that we will perform. On the one hand, the homogeneous
sector has been described in the Ashtekar formulation
in order to prepare it for the polymeric quantization
characteristic of LQC. On the other hand, for the
inhomogeneous sector we have chosen the
parametrization introduced in Ref. \cite{men1},
since it is essentially the only one which admits a
satisfactory Fock quantization, including a unitary
implementation of the dynamics after completing the
choice of time gauge (i.e. after a complete
deparametrization). The coupling between the two
sectors shows the interest (and nontriviality) of our
approach because, owing to it, it is not
straightforward that a well-defined hybrid
quantization be viable.

\section{Fock Quantization}
\setcounter{subsection}{0}

We will start the construction of our quantum model
with the Fock quantization of the inhomogeneous sector
and the imposition of the constraint defined by the
generator of $S^1$-translations, which depends only on
the inhomogeneities. In doing so, we first choose as
complex structure the one which is naturally
associated with the identification of
$\{(a_m,a_m^*)\}$ as creation and annihilation
operators, for all modes with $m\neq 0$ \cite{men2}.
Using this complex structure, we construct the
``one-particle'' Hilbert space, the corresponding
symmetric Fock space $\mathcal F$, and a quantization
of the variables $\{(a_m,a_m^*)\}$ by standard
methods. The field dynamics is unitarily implemented
in this quantization \cite{men1}. Besides, the
quantization provides also a natural unitary
implementation of the gauge group of
$S^1$-translations, since the resulting vacuum is
invariant under that group. Imposing these two
unitarity requirements, the Fock quantization turns
out to be unique (up to equivalence) \cite{men2}.

We will denote by $|\{n_m\}\rangle:=
|...,n_{-m},...,n_m,...\rangle$ the corresponding
$n$-particle states, which provide an orthonormal
basis for the Fock space. Here $n_m<\infty$ is the
occupation number of the $m$-th mode. In these states,
only a finite set of these occupation numbers differ
from zero. The dense set spanned by them will be
denoted by $\mathcal S$.

The generator of $S^1$-translations, given in Eq.
\eqref{classicCtheta}, can be promoted to the quantum
operator:
\begin{equation}\label{quantumCtheta}
\widehat C_\theta=\sum_{m=1}^\infty m (\hat
a^\dagger_m \hat a_m-\hat a^\dagger_{-m} \hat a_{-m}).
\end{equation}
The proper Fock subspace annihilated by this
constraint will be called $\mathcal F_p$. A basis for
it is provided by the $n$-particle states which verify
the condition
\begin{equation}\label{S1symmetry}
\sum_{m=1}^\infty m(n_m-n_{-m})=0.
\end{equation}
Clearly, the vacuum is one of these states.

The subspace $\mathcal F_p$ is the physical Hilbert
space of the deparametrized system quantized in Ref.
\cite{men1}. There, $\widehat C_\theta$ was the
only quantum constraint. In order to get the physical
Hilbert space now, however, we need to impose also the
quantum counterpart of the Hamiltonian constraint
\eqref{classicCG}.

\section{Quantization of the Bianchi I Model}
\setcounter{subsection}{0}

When the inhomogeneities vanish, the Gowdy model
reduces to the vacuum Bianchi I model with $T^3$
topology. As a preliminary step before discussing the
quantization of the complete Gowdy system, including
both the homogeneous and inhomogenous sectors, we will
focus our attention on the Bianchi I model, quantizing
it polymerically. Here, we will summarize and revisit
the analysis of Ref. \cite{mmp}, where we
accomplished the loop quantization of this particular
(homogeneous) subfamily of Gowdy spacetimes.

\subsection{Quantum representation}

In LQC the basic configuration variables are
holonomies, whereas the basic momentum variables are
fluxes. The holonomy along an edge of oriented
coordinate length $2\pi\mu_i$ in the direction $i$
is $h_i^{\mu_i}(c^i)=e^{
\mu_{i}c^{i}\tau_{i}}$ \cite{note}, where $\tau_i$ are the $SU(2)$
generators proportional to the Pauli matrices, such
that $[\tau_i,\tau_j]=\epsilon_{ijk}\tau^k$. On the
other hand, the flux across a square $S^i$ of fiducial
area $(2\pi\mu_i)^2$ and normal to the direction $i$
is $E(S^i)=p_i\,(\mu_i)^2$.

The configuration algebra for each fiducial direction
$i$ is the algebra of almost periodic functions of
$c^i$, which is generated by the matrix elements of
the holonomies $\mathcal
N_{\mu_i}(c^i)=e^{\frac{i}{2}\mu_{i}c^{i}}$. We will
call $\text{Cyl}_\text{S}^i$ the corresponding vector
space. The kinematical Hilbert space $\mathcal
H_{\text{kin}}$ is the tensor product of three copies
$\mathcal H_{\text{kin}}^i$ (one for each fiducial
direction) of the space $L^2(\mathbb{R}_{\text{B}},
d\mu_{\text{Bohr}})$, where $\mathbb{R}_{\text{B}}$ is
the Bohr compactification of the real line, and
$d\mu_{\text{Bohr}}$ is the normalized Haar measure on
it \cite{vel}. In momentum representation, and
employing the Dirac ket notation $|\mu_i\rangle$ to
denote the states $\mathcal N_{\mu_i}(c^i)$, the
Hilbert space $\mathcal H_{\text{kin}}^i$ can be seen
as the completion of the algebra
$\text{Cyl}_\text{S}^i$ with respect to the discrete
inner product
$\langle\mu_i|\mu_i^\prime\rangle=\delta_{\mu_i
\mu_i^\prime}$.

On the basis states $|\mu_i\rangle$, the action of the
basic operators  $\hat p_i$ (associated with fluxes)
and $\hat{\mathcal N}_{\mu_i^\prime}$ (associated with
holonomies) reads
\begin{equation}\label{action}
\hat p_i|\mu_i\rangle=4\pi\gamma
l_\text{Pl}^2\mu_i|\mu_i\rangle,
\qquad
\hat{\mathcal N}_{\mu_i^\prime}|\mu_i\rangle=|\mu_i
+\mu_i^\prime\rangle.
\end{equation}
Here, $l_\text{Pl}=\sqrt{G\hbar}$ is the Planck length.

\subsection{Hamiltonian constraint}

The classical Hamiltonian constraint
\eqref{classicCBI} of the Bianchi I model leads to the
following form when expressed in terms of the basic
variables of LQC:
\begin{equation}\label{classicCBIholonomy}
C_{BI}=-\frac{2}{\gamma^2} \left[\Lambda_\theta
\Lambda_\sigma\left(\frac{1}{\sqrt{|p_\delta|}}\right)
+\Lambda_\theta\Lambda_\delta
\left(\frac{1}{\sqrt{|p_\sigma|}}\right)+
\Lambda_\sigma\Lambda_\delta \left(\frac{1}
{\sqrt{|p_\theta|}}\right)\right],
\end{equation}
where
\begin{eqnarray}\label{Lambda}
\Lambda_i=-\lim_{{\mu_i^\prime}\rightarrow0}
\frac{i\sqrt{|p_i|}}{2{\mu_i^\prime}}\text{sign}(p_i)
({\mathcal N}_{2\mu_i^\prime}-{\mathcal
N}_{-2\mu_i^\prime}).
\end{eqnarray}

The existence in Loop Quantum Gravity of a minimum
nonzero eigenvalue for the area,
$\Delta=2\sqrt{3}\pi\gamma l_{\text{Pl}}^2$, has been
argued to imply that the limit
$\mu_i^\prime\rightarrow0$ is not feasible and that,
for each fiducial direction, there exists in fact a
minimum nonzero edge length for the holonomy,
$\bar\mu_i$. Therefore, in order to obtain the quantum
counterpart of $\Lambda_i$ in LQC, the above classical
limit is replaced by evaluation at $\bar\mu_i$. On the
other hand, to determine this minimum fiducial length
we adopt the proposal presented in Ref.
\cite{chio}, which leads to the condition
$\bar{\mu}_i^2 |p_i|=\Delta$. Thus, the value of
$\bar{\mu}_i$ depends on that of $|p_i|$, which is one
of the degrees of freedom of the physical metric.
Quantum mechanically, this translates into the
operator relation
\begin{equation}\label{muop}
\widehat{\frac{1}{\bar\mu_i}}=
\frac{\widehat{\sqrt{|p_i|}}}{\sqrt{\Delta}}.
\end{equation}
Taking then a suitable symmetric factor ordering in
the quantum counterpart of \eqref{Lambda}, we finally
arrive at the operator
\begin{equation}\label{Lambdasym}
\widehat{\Lambda}_i=-\frac{i}{4\sqrt{\Delta}}
\widehat{\sqrt{|p_i|}} \bigg[(\hat{\mathcal
N}_{2\bar\mu_i}-\hat{\mathcal N}_{-2\bar\mu_i})
\widehat{\text{sign}(p_i)}
+\widehat{\text{sign}(p_i)}(\hat{\mathcal
N}_{2\bar\mu_i}-\hat{\mathcal N}_{-2\bar\mu_i})\bigg]
\widehat{\sqrt{|p_i|}}.
\end{equation}

In principle, given relation \eqref{muop}, the
operator $\hat{\mathcal N}_{\bar\mu_i}$ should produce
a state-dependent shift $\bar\mu_i(\mu_i)$ on the
basis states $|\mu_i\rangle$. It is then most
convenient to relabel this basis introducing an affine
parameter $v_i$ such that the action of $\hat{\mathcal
N}_{\bar\mu_i}$ is defined to cause just a constant
shift in the new label. Such an affine
reparametrization is possible (see e.g. Ref.
\cite{mmp}), resulting in the following action of
the basic operators on the states $|v_i\rangle$ of the
relabeled basis:
\begin{equation}
\hat p_i|v_i\rangle=3^{1/3}\Delta\,\text{sign}(v_i)
|v_i|^{2/3}|v_i\rangle,\qquad\hat{\mathcal
N}_{\bar\mu_i}|v_i\rangle=|v_i+1\rangle.
\end{equation}

Starting with $\hat p_i$, we can define the operators
$\widehat{\sqrt{|p_i|}}$ and
$\widehat{\text{sign}(p_i)}$ via the spectral
theorem \cite{spectral}. It is then straightforward to
compute the action on the above basis of the operator
$\widehat{\Lambda}_i$, given in Eq. \eqref{Lambdasym}.
To promote the Hamiltonian constraint
\eqref{classicCBIholonomy} to an operator, we also
need to represent the inverse of $\sqrt{|p_i|}$. In
this case, the spectral theorem is not applicable
since zero is in the discrete spectrum of $\hat p_i$.
To overcome this problem, one introduces the quantum
analog of the classical identity \cite{note}
\begin{align}\label{reg}
\left(\frac{1}{\sqrt{|p_i|}}\right)&=
\frac{\text{sign}(p_i)}{2\pi
G\bar\mu_i}\text{tr}\left(\tau_{i}h_i^{\bar\mu_i}(c^i)
\left\{h_i^{-\bar\mu_i}(c^i),
\sqrt{|p_i|}\right\}\right).
\end{align}
The result is a regularized operator whose action on
the basis states turns out to be diagonal:
\begin{equation}\label{inv}
\widehat{\left[\frac{1}{\sqrt{|p_i|}}\right]}
|v_i\rangle=\frac{b(v_i)}{\sqrt{\gamma}l_{\text{Pl}}}
|v_i\rangle,\quad b(v_i)=\frac{3^{7/12}}{\sqrt{8\pi}}
|v_i|^{1/3}\left||v_i+1|^{1/3}-|v_i-1|^{1/3}\right|.
\end{equation}

In this way, the Hamiltonian constraint of the Bianchi
I model is represented by the symmetric operator
\begin{equation}\label{quantumCBI}
\widehat C_{BI}=-\frac{2}{\gamma^2}
\left\{\hat\Lambda_\theta\hat\Lambda_\sigma
\widehat{\left[\frac{1}{\sqrt{|p_\delta|}}\right]}
+\hat\Lambda_\theta\hat\Lambda_\delta
\widehat{\left[\frac{1}{\sqrt{|p_\sigma|}}\right]}+
\hat\Lambda_\sigma\hat\Lambda_\delta
\widehat{\left[\frac{1}{\sqrt{|p_\theta|}}\right]}\right\},
\end{equation}
which is densely defined on the domain
$\text{Cyl}_\text{S}=\text{span}
\{|v_\theta,v_\sigma,v_\delta\rangle= |v_\theta
\rangle\otimes |v_\sigma\rangle \otimes
|v_\delta\rangle\}$.

\subsection{Singularity resolution}

Owing to the factor ordering chosen in our
definitions, our symmetric Hamiltonian constraint
annihilates all the states in the basis
$\{|v_\theta,v_\sigma,v_\delta\rangle\}$ with any of
the $v_i$'s equal to zero. We will call them ``zero
volume states'', since they are eigenstates of the
(homogeneous-)volume operator $\widehat
V=\otimes_i\widehat{\sqrt{|p_i|}}$ with vanishing
eigenvalue. Furthermore, the complement of the
subspace spanned by these zero volume states is
invariant under the action of the constraint operator
\eqref{quantumCBI}. Therefore, the restriction of the
constraint to this complement is well defined. We can
say that, when the constraint is imposed, this
complement decouples from the space of zero volume
states. Then, we limit our considerations to it in the
following in order to find nontrivial solutions to the
constraint. We will call $\widetilde{\text{Cyl}}_S$
the linear span of tensor products of states
$|v_i\rangle$ such that none of the $v_i$'s vanishes,
whereas $\widetilde{\mathcal H}_\text{Kin}$ will
denote the corresponding Hilbert space of nonzero
volume states.

It is worth noticing that, since zero volume states
have been removed from our kinematical Hilbert space,
so that the kernel of all the operators $\hat{p}_i$ is
empty, there is no longer any quantum analog of the
classical cosmological singularity, where some of the
triads $p_i$ vanish classically. In this sense, the
singularity is resolved in our Bianchi I model. We
will discuss this issue in more detail in Sect. 6,
where we will also analyze the fate of the singularity
in the Gowdy model.

\subsection{Densitized Hamiltonian constraint}
\setcounter{subsubsection}{0}

In principle, solutions to the Hamiltonian constraint
do not need to be normalizable in the kinematical
Hilbert space, which is mainly a mathematical tool
introduced to construct a representation of the
holonomy-flux algebra and the constraints of the
model, but which does not take into account the
dynamics. We will look for solutions in the much
bigger space $(\widetilde{\text{Cyl}}_S)^*$, the
algebraic dual of the dense domain of definition of
the Hamiltonian constraint, whose elements will be
denoted by $(\psi|$. Instead of looking directly for
solutions to the constraint \eqref{quantumCBI}, which
are difficult to determine, we find more convenient to
densitize this constraint by means of the following
bijection in the dual space
$(\widetilde{\text{Cyl}}_S)^*$:
\begin{equation}\label{map}
(\psi|\longrightarrow(\psi|
\widehat{\left[\frac{1}{V}\right]}^{\frac1{2}},
\end{equation}
where the operator that represents the inverse of the
(homogeneous-)volume is
\begin{equation}\label{vol}
\widehat{\left[\frac{1}{V}\right]}=
\otimes_i\widehat{\left[\frac{1}{\sqrt{|p_i|}}\right]}.
\end{equation}

The transformed physical states are now annihilated by
the (adjoint of the) symmetric densitized Hamiltonian
constraint, defined as
\begin{equation}\label{densitizedCBI}
\widehat{{\cal
C}}_{\text{BI}}=\widehat{\left[\frac{1}{V}
\right]}^{-\frac1{2}}\widehat{C}_{\text{BI}}
\widehat{\left[\frac{1}{V}
\right]}^{-\frac1{2}}=
-\frac{2}{\gamma^2}\bigg[\widehat{\Theta}_\theta
\widehat{\Theta}_\sigma+\widehat{\Theta}_\theta
\widehat{\Theta}_\delta+
\widehat{\Theta}_\sigma\widehat{\Theta}_\delta\bigg],
\end{equation}
where $\widehat{\Theta}_i$ is the symmetric operator
\begin{equation}
\widehat{\Theta}_i=
\widehat{\left[\frac{1}{\sqrt{|p_i|}}
\right]}^{-\frac1{2}}\widehat{\Lambda}_i
\widehat{\left[\frac1{\sqrt{|p_i|}}
\right]}^{-\frac1{2}}.
\end{equation}
Note that the operator
$\widehat{[1/\sqrt{|p_i|}]}{}^{-1/2}$ is well defined
once we have restricted the study to (a dense domain
in) the space of nonzero volume states. It is also
worth noticing that the above kind of densitization
procedure can be similarly applied if
$\widetilde{\text{Cyl}}_S$ is replaced with another
dense domain for the Hamiltonian constraint such that
the inverse (homogeneous)-volume is still a bijection
in the corresponding dual.

From Eq. \eqref{densitizedCBI}, we see that all the
operators $\widehat{\Theta}_i$ are Dirac observables
in the Bianchi I model, because they commute with the
constraint. This fact simplifies enormously the
resolution of the constraint, which can be seen as an
algebraic equation in the eigenstates that are allowed
simultaneously for the three operators
$\widehat{\Theta}_i$, one for each direction. Since
the three operators are formally identical, the
problem is reduced to a one-dimensional problem. With
this motivation, we will now analyze the properties of
$\widehat{\Theta}_i$.

\subsubsection{Superselection and no-boundary description}

As one can check by direct calculation, the operator
$\widehat{\Theta}_i$ is a difference operator acting
on the basis states $|v_i\rangle$:
\begin{equation}\label{acttheta}
\widehat{\Theta}_i|v_i\rangle=-i\frac{\Delta}{2\sqrt{3}}
\big[f_+(v_i)|v_i+2\rangle-f_-(v_i)|v_i-2\rangle\big],
\end{equation}
where
\begin{equation}\label{f}
f_\pm(v_i)=g(v_i\pm2)s_\pm(v_i)g(v_i), \qquad
s_\pm(v_i)=\text{sign}(v_i\pm2)+\text{sign}(v_i),
\end{equation}
and
\begin{align}\label{g}
g(v_i)&=
\begin{cases}
\left|\left|1+\frac1{v_i}\right|^{\frac1{3}}
-\left|1-\frac1{v_i}\right| ^{\frac1{3}}
\right|^{-\frac1{2}} & {\text{if}} \quad v_i\neq 0,\\
0 & {\text{if}} \quad v_i=0.\\
\end{cases}
\end{align}
The presence of signs in Eq. \eqref{f} has important
consequences. Namely, it turns out that the function
$f_+(v_i)$ ($f_-(v_i)$) vanishes in the whole interval
$[-2,0]$ ($[0,2]$). As a result, in terms of the label
$v_i$, the action of the difference operator
$\widehat{\Theta}_i$ does not mix any of the following
semilattices:
\begin{equation}
\mathcal
L_{\varepsilon_i}^\pm=\{\pm(\varepsilon_i+2k),
k=0,1,2...\},\qquad\varepsilon_i\in(0,2].
\end{equation}
From now on, we will call $\mathcal
H_{\varepsilon_i}^{\pm}$ the corresponding subspaces
of states with support in these semilattices. Then,
each of them provides a superselection sector.

If we compare the classical densitized constraint [see
Eqs. \eqref{classicCG} and \eqref{classicCBI}] with
its quantum counterpart, given in Eq.
\eqref{densitizedCBI}, we conclude that the operator
$\widehat{\Theta}_i$ is the quantum analog of the
classical quantity $c^ip_i$ \cite{note}. Hence, the
Wheeler-DeWitt analog of $\widehat{\Theta}_i$ would be
the first order differential operator
$\widehat{\underline{\Theta}}_i=i8\pi \gamma
l_{\text{Pl}}^2p_i\partial_{p_i}$. One would expect
that $\widehat{\underline{\Theta}}_i$ could be
recovered from $\widehat{\Theta}_i$ in a suitable
semiclassical limit. However, its action
\eqref{acttheta} indicates that $\widehat{\Theta}_i$
is instead a second-order difference operator. This
apparent conflict, nonetheless, does not break
physical consistency. Because of the property
$f_{\mp}(\varepsilon_i)=0$, when one hits the origin
in $\mathcal H_{\varepsilon_i}^{\pm}$, one gets a
relation which constraints the data at
$v_i=\pm\varepsilon_i$ and $v_i=\pm(\varepsilon_i+2)$.
Hence, the eigenstates of $\widehat{\Theta}_i$ in
$\mathcal H_{\varepsilon_i}^{\pm}$ depend in fact on a
single piece of initial data (as it would correspond
to a first-order operator), given just by the
projection on the $v_i=\pm\varepsilon_i$ ``slice'' of
the semilattice.

In this sense, besides, the Hamiltonian constraint
provides a no-boundary description: physical states
will not only have no contribution from the slice
$v_i=0$ (for any $i$=1, 2 or 3), which would
correspond to the classical singularity, but they also
get no contribution with the opposite sign of the
label $v_i$ --namely the opposite triad orientation--
without the need to impose any boundary condition at
the initial slice.

\subsubsection{Spectrum and eigenfunctions}
\label{subsec:spe}

Let us call $\text{Cyl}_{\varepsilon_i}^{\pm}$ the
linear span of the $v_i$-states in the semilattice
$\mathcal L_{\varepsilon_i}^\pm$. Then, the operator
$\widehat{\Theta}_i$, with domain
$\text{Cyl}_{\varepsilon_i}^{\pm}$, is essentially
self-adjoint on $\mathcal
H_{\varepsilon_i}^{\pm}$ \cite{mmp}. In addition, its
spectrum has been completely characterized: it is
absolutely continuous, nondegenerate, and coincides
with the real line. Furthermore, the generalized
eigenfunction with generalized eigenvalue
$\lambda\gamma l_{\text{Pl}}^2$, denoted by
$e^{\pm\varepsilon_i}_{\lambda}(v_i)$, is completely
determined by the data on the initial slice,
$e^{\pm\varepsilon_i}_{\lambda}(\pm\varepsilon_i)$, as
we pointed out before. One can show that the explicit
expression e.g. of $e^{\varepsilon_i}_{\lambda}(v_i)$
is \cite{mmp}
\begin{equation}\label{eigen}
e^{\varepsilon_i}_{\lambda}(\varepsilon_i+2M)=
\sum_{O(M)} \left[\prod_{\{r_k\}}
\frac{f_-(\varepsilon_i+2r_k+2)}
{f_+(\varepsilon_i+2r_k+2)}\right]
\left[\prod_{\{s_l\}} \frac{-i\,2\sqrt{3}\lambda
\gamma l_{\text{Pl}}^2}{\Delta\,
f_+(\varepsilon_i+2s_l)}\right]
e^{\varepsilon_i}_{\lambda}(\varepsilon_i).
\end{equation}
Here, $O(M)$ denotes the set of all possible ways to
move from 0 to $M$ by jumps of one or two steps. For
each element in $O(M)$, $\{r_k\}$ is the subset of
integers followed by a jump of two steps, whereas
$\{s_l\}$ is the subset of integers followed by a jump
of only one step. Note that, up to a constant phase,
these complex coefficients oscillate from real to
imaginary when $v_i$ varies along the considered
semilattice.

After a suitable (delta-)normalization of the
generalized eigenstates
$|e_\lambda^{\pm\varepsilon_i}\rangle$ (in ket
notation), the spectral resolution of the identity in
the kinematical Hilbert space $\mathcal
H_{\varepsilon_i}^{\pm}$ associated with
$\widehat{\Theta}_i$ is given then by
\begin{equation}\label{identity}
\mathbb{I}^\pm_{\varepsilon_i}=\int_{\mathbb{R}}
d\lambda |e_{\lambda}^{^\pm\varepsilon_i}\rangle
\langle e_{\lambda}^{^\pm\varepsilon_i}|.
\end{equation}

\subsection{Physical states for Bianchi I}
\setcounter{subsubsection}{0}

Let us restrict our study to any specific
superselection sector determined by the three numbers
$(\varepsilon_\theta,
\varepsilon_\sigma,\varepsilon_\delta)$ and by a sign
in each direction for the orientation of the triad. In
the following, for simplicity, we will choose positive
orientations. Since the $\widehat{\Theta}_i$'s are
Dirac observables and we know their associated
resolution of the identity, it is really easy to find
the physical Hilbert space. We can follow two
strategies. On the one hand, since
$\widehat{\Theta}_i$ is essentially self-adjoint in
$\text{Cyl}_{\varepsilon_i}^{+}$, the constraint
operator $\widehat{\mathcal{C}}_{\text{BI}}$, given in
Eq. \eqref{densitizedCBI}, is essentially self-adjoint
in the tensor product of these spaces, and we can
apply the group averaging procedure \cite{gave,gave2}
to determine the solutions to the Hamiltonian
constraint and their Hilbert structure. On the other
hand, we can also solve directly the constraint and
determine the Hilbert structure of the solutions by
choosing a complete set of real observables and
imposing that they be represented as self-adjoint
operators.

In both cases, one concludes that the solutions to the
Hamiltonian constraint have the form
\begin{equation}\label{phystates}
\psi(v_\theta,v_\sigma,v_\delta)=
\int_{\mathbb{R}^2}d\lambda_\sigma d\lambda_\delta
e_{\lambda_\theta[\lambda]}^{\varepsilon_\theta}(v_\theta)
e_{\lambda_\sigma}^{\varepsilon_\sigma}(v_\sigma)
e_{\lambda_\delta}^{\varepsilon_\delta}(v_\delta)
\tilde\psi(\lambda_\sigma,\lambda_\delta),
\end{equation}
with
\begin{equation}\label{lambda}
\lambda_\theta[\lambda]=
-\frac{\lambda_\delta\lambda_\sigma}
{\lambda_\delta+\lambda_\sigma}.
\end{equation}
Here, the physical states $\tilde\psi(\lambda_\sigma,
\lambda_\delta)$ belong to the Hilbert space
\begin{equation}\label{physpaceBI}
\mathcal H^{\text{BI}}=L^2\left(\mathbb{R}^2,
|\lambda_\sigma+\lambda_\delta|d\lambda_\sigma
d\lambda_\delta\right).
\end{equation}

\subsubsection{Evolution and observables}
\label{subsec:obs}

In the above expression for the solutions, we have
eliminated the dependence on $\lambda_\theta$,
determined in terms of $\lambda_\sigma$ and
$\lambda_\delta$. This is just a matter of convention:
we could choose to fix any of these three variables in
terms of the other two. In view of our choice, we can
interpret $v_\theta$ as an internal time and regard
physical states as evolving with respect to
it \cite{mmp2}. At a particular time $v_\theta^o$, the
solution can be written as
\begin{equation}\label{phystates}
\psi(v_\sigma,v_\delta)|_{v_\theta^o}=
\int_{\mathbb{R}^2}d\lambda_\sigma d\lambda_\delta
e_{\lambda_\sigma}^{\varepsilon_\sigma}(v_\sigma)
e_{\lambda_\delta}^{\varepsilon_\delta}(v_\delta)
\tilde\psi(\lambda_\sigma,\lambda_\delta)|_{v_\theta^o},
\end{equation}
where
\begin{equation}
\tilde\psi(\lambda_\sigma,\lambda_\delta)|_{v_\theta^o}=
e_{\lambda_\theta[\lambda]}^{\varepsilon_\theta}
(v_\theta^o)\tilde\psi(\lambda_\sigma,\lambda_\delta)
\end{equation}
belongs to the ``$v_\theta^o$-slice'' Hilbert space
\begin{equation}\label{vspaceBI}
\mathcal H_{v_\theta^o}=L^2\left(\mathbb{R}^2,
\frac{|\lambda_\sigma+\lambda_\delta|}
{\big|e_{\lambda_\theta[\lambda]}^{\varepsilon_\theta}
(v_\theta^o)\big|^{2}}d\lambda_\sigma
d\lambda_\delta\right).
\end{equation}

A complete set of observables is given by the
constants of motion $\widehat{\Theta}_\delta$ and
$\widehat{\Theta}_\sigma$, which act on the physical
states just by multiplication by $\lambda_\sigma\gamma
l_{\text{Pl}}^2$ and $\lambda_\delta\gamma
l_{\text{Pl}}^2$ respectively, and by the observables
at fixed time $\hat v_\sigma|_{v_\theta^o}$ and $\hat
v_\delta|_{v_\theta^o}$, whose action on solutions is
given by
\begin{equation}
\hat v_\alpha|_{v_\theta^o}\,\psi(v_\sigma,v_\delta)
|_{v_\theta^o}=v_\alpha\psi(v_\sigma,v_\delta)|_{v_\theta^o},
\quad\quad \alpha=\sigma,\delta.\end{equation} On the
$v_\theta^o$-slice Hilbert space, on the other hand,
the corresponding action is
\begin{equation}
\hat v_\sigma|_{v_\theta^o}\tilde\psi(\lambda_\sigma,
\lambda_\delta)|_{v_\theta^o}=
\int_{\mathbb{R}}d\tilde\lambda_\sigma \langle
e_{\lambda_\sigma}^{\varepsilon_\sigma}|v_\sigma
|e_{\tilde\lambda_\sigma}^{\varepsilon_\sigma}
\rangle_{_{\text{kin}}}
\tilde\psi(\tilde\lambda_\sigma,
\lambda_\delta)|_{v_\theta^o},
\end{equation}
and similarly for $\hat v_\delta|_{v_\theta^o}$.

The fact that the dependence of $|e_{\lambda_\theta
[\lambda]}^{\varepsilon_\theta}(v_\theta)|$ on
$\lambda_\sigma$ and $\lambda_\delta$ varies with the
value of $v_\theta$ precludes one establishing a
unitary relation between the different
$v_\theta$-slice Hilbert spaces, obtained for other
choices of $v_\theta$ instead of $v_\theta^o$, so that
one does not get a unitary evolution in this emergent
time. This result is a mere consequence of the nature
of the polymeric quantization performed on the
variable that plays the role of internal
time \cite{mmp2}.

\section{Hybrid Quantization of the Gowdy Model}
\setcounter{subsection}{0}

Once we have represented the homogeneous sector of the
Gowdy model following the prescriptions of LQC, and
the inhomogeneous sector employing the Fock
quantization, we are ready to construct the quantum
counterpart of the Hamiltonian constraint of the Gowdy
model, $C_{\text{G}}$, within our hybrid approach. The
kinematical Hilbert space is just the tensor product
of the kinematical Hilbert spaces for each sector,
that is $\mathcal H_{\text{kin}}\otimes\mathcal F$.

Taking into account Eqs.
\eqref{classicCG}-\eqref{Hamiltonians}, it is
straightforward to represent the constraint
$C_{\text{G}}$ as an operator. In fact, in the
previous section we have already represented the
quantum analog $\widehat C_{\text{BI}}$ of the
homogeneous term $C_{\text{BI}}$. In addition, we have
constructed the inverse homogeneous-volume operator,
given in Eq. \eqref{vol}, and obtained the quantum
counterpart $\widehat{\Theta}_i$ of the quantity
$c^ip_i$. Therefore, we can also promote the
inhomogeneous term $C_\xi$ to an operator. Adopting
the same kind of factor ordering used for the
homogeneous term, we get the following symmetric
operator:
\begin{equation}\label{quantumCXI}
\widehat{C}_\xi=\widehat{\left[\frac1{V}
\right]}^{\frac1{2}}\widehat{\mathcal
C}_\xi\widehat{\left[\frac{1}{V}
\right]}^{\frac1{2}},\quad
\widehat{\mathcal{C}}_{\xi}=
l_{\text{Pl}}^2\bigg[\frac{( \widehat{\Theta}_\sigma
+\widehat{\Theta}_\delta)^2}
{\gamma^2}\bigg(\widehat{\frac{1}
{\sqrt{|p_\theta|}}}\bigg)^2
\widehat{H}_\text{int}^\xi+32\pi^2
\widehat{|p_\theta|}\widehat{H}_0^\xi\bigg],
\end{equation}
where the free Hamiltonian $\widehat{H}_0^\xi$ and the
interaction term $\widehat{H}_\text{int}^\xi$ are
normal ordered:
\begin{equation}\label{quantumHamiltonians}
\widehat{H}_0^\xi=\sum_{m\neq0}|m|\hat{a}^{\dagger}_m
\hat{a}_m,\qquad
\widehat{H}_\text{int}^\xi=\sum_{m\neq0}
\frac{1}{2|m|}(2\hat{a}^{\dagger}_m
\hat{a}_m+ \hat{a}_m\hat{a}_{-m}+
\hat{a}^{\dagger}_m\hat{a}^{\dagger}_{-m}).
\end{equation}

Similarly to what happens with the Hamiltonian
constraint of Bianchi I, $\widehat C_{\text{BI}}$, the
inhomogeneous term $\widehat{C}_\xi$ annihilates the
subspace spanned by the zero homogeneous-volume states
and leaves invariant its complement. Therefore, this
subspace decouples from its complement also in the
complete Gowdy model when the constraint is imposed.
Thus, we can again restrict our study to the
kinematical Hilbert space $\widetilde{\mathcal
H}_{\text{kin}}$ in the homogeneous sector. Moreover,
the Hilbert spaces $\otimes_i\mathcal
H_{\varepsilon_i}^{\pm}\otimes\mathcal F$ are also
superselected by the Hamiltonian constraint of the
Gowdy model, so that for practical purposes we can
further restrict our discussion to any of these
kinematical Hilbert spaces. We continue choosing
positive orientation of the triads for simplicity and,
in principle, define our operators in the
corresponding dense set
\begin{equation}
\otimes_i\text{Cyl}_{\varepsilon_i}^{+}\otimes
\mathcal
S=\text{span}\{|v_\theta,v_\sigma,v_\delta\rangle
\otimes|\{n_m\}\rangle; \quad v_i\in\mathcal
L_{\varepsilon_i}^+\}.
\end{equation}

As in the Bianchi I case, it is preferable to
densitize the Hamiltonian constraint, in particular
because it is then straightforward to recognize some
Dirac observables, what facilitates the resolution of
the constraint. We introduce again the bijective map
\eqref{map} [in principle in the dual
$(\otimes_i\text{Cyl}_{\varepsilon_i}^{+}\otimes\mathcal
S)^*$, but the dual of any other dense set where the
inverse homogenous-volume operator provides a
bijection would be acceptable as well]. This leads to
``transformed'' physical states which are annihilated
just by the densitized Hamiltonian constraint
$\widehat{\mathcal C}_{\text{G}}=\widehat {\mathcal
C}_{\text{BI}}+\widehat{\mathcal C}_\xi$, with
$\widehat {\mathcal C}_{\text{BI}}$ given in Eq.
\eqref{quantumCBI} and $\widehat{\mathcal C}_\xi$ in
Eq. \eqref{quantumCXI}. It is worth noting that,
although the constraint operator $\widehat{\mathcal
C}_{\text{G}}$ couples both the homogenous and
inhomogeneous sectors of the Gowdy model in a highly
nontrivial way, it is indeed a well-defined symmetric
operator with domain
$\otimes_i\text{Cyl}_{\varepsilon_i}^{+}\otimes\mathcal
S$.

In the Gowdy model, the operators
$\widehat{\Theta}_\sigma$ and
$\widehat{\Theta}_\delta$ are still Dirac observables.
Nevertheless, the densitized Hamiltonian constraint
depends now on $\widehat{|p_\theta|}$ and
$\widehat{1/\sqrt{|p_\theta|}}$, so that
$\widehat{\Theta}_\theta$ no longer commutes with this
constraint and fails to be a Dirac observable. It is
then convenient to use the basis of states
$|v_\theta\rangle\otimes
|e_{\lambda_\sigma}^{\varepsilon_\sigma}\rangle
\otimes|e_{\lambda_\delta}^{\varepsilon_\delta}
\rangle$ for the homogeneous sector. In doing so, the
densitized constraint becomes a difference equation in
$v_\theta$ and can be regarded as an evolution
equation in this parameter, what means that $p_\theta$
plays the role of internal time. As we have already
commented, this was essentially the time choice
adopted to deparametrize the system in the Fock
quantization of Ref. \cite{men1}. Apart from a
global factor and some suitable rescalings, the
inhomogeneous part of the constraint,
$\widehat{\mathcal C}_\xi$, coincides in each
``generalized eigenspace'' of
$\widehat{\Theta}_\sigma$ and
$\widehat{\Theta}_\delta$ with the inhomogeneous
Hamiltonian of the deparametrized (gauge-fixed)
model \cite{men1}. The time dependence of that
Hamiltonian becomes a dependence on $p_\theta$ in our
case.

\subsection{Solutions to the eigenvalue equation for
the Hamiltonian constraint}

Let us consider the (complex) eigenvalue equation for
$\widehat{\mathcal C}_{\text{G}}$,
\begin{equation}
(\psi|\widehat{\mathcal C}_{\text{G}}= \rho
l_{\text{Pl}}^4(\psi|.
\end{equation}
Using the basis of states introduced above for the
homogeneous sector and substituting the \emph{formal}
expansion
\begin{equation}
(\psi|=\sum_{v_\theta\in\mathcal
L_{\varepsilon_\theta}^+}\int_{\mathbb{R}^2}
d\lambda_\sigma d\lambda_\delta \langle
v_\theta|\otimes \langle
e_{\lambda_\sigma}^{\varepsilon_\sigma}|
\otimes\langle
e_{\lambda_\delta}^{\varepsilon_\delta}| \otimes
(\psi_{\lambda_\sigma,\lambda_\delta} (v_\theta)|,
\end{equation}
we get the solution \cite{let,new}
\begin{eqnarray}\label{solution}
\big(\psi_{\lambda_\sigma,\lambda_\delta}
(\varepsilon_\theta+2M)\big|\{n_m\}\big
\rangle&=&\big(\psi_{\lambda_\sigma,\lambda_\delta}
(\varepsilon_\theta)\big|\sum_{O(M)}
\left[\prod_{\{r_k\}}\frac{f_-(\varepsilon_\theta+2r_k+2)}
{f_+(\varepsilon_\theta+2r_k+2)}\right]\nonumber
\\&\times&\mathcal
P\Big[\prod_{\{s_l\}}\widehat{H}^\xi_\rho
(\varepsilon_\theta+2s_l,\lambda_\sigma,
\lambda_\delta)\Big]\,\big|\{n_m\}\big\rangle,
\end{eqnarray}
where $O(M)$, $\{r_k\}$ and $\{s_l\}$ have the same
meaning as in Sec. \eqref{subsec:spe}. Besides, the
symbol $\mathcal P$ denotes path ordering, and
$\widehat{H}^\xi_\rho(v_\theta,
\lambda_\sigma,\lambda_\delta)$ has the form:
\begin{eqnarray}
\widehat{H}^\xi_\rho(v_\theta,
\lambda_\sigma,\lambda_\delta)&=&
\frac{i}{2\pi(\lambda_\sigma+\lambda_\delta)
f_+(v_\theta)}\bigg[\rho+2\lambda_\sigma
\lambda_\delta\nonumber\\
&-&\frac{(\lambda_\sigma+\lambda_\delta)^2}
{\gamma}b^2(v_\theta) \widehat H_{\text{int}}^\xi- 32
\pi^2  3^{1/3}\gamma \Delta |v_\theta|^{2/3}\widehat
H_0^\xi\bigg].
\end{eqnarray}

\subsection{Physical states and observables}

The operator $\widehat{\mathcal C}_{\text{G}}$ is
essentially self-adjoint if and only if there is no
normalizable solution of the form \eqref{solution}
when $\rho=\pm i$. In principle, one can make use in
that case of the group averaging
method \cite{gave,gave2} to construct the physical
Hilbert space. In practice, nevertheless, to apply
this method we should first determine and adopt an
invariant domain for the constraint operator
$\widehat{\mathcal C}_{\text{G}}$, a task which seems
difficult to accomplish given the complexity of the
action of the constraint and the infinite number of
degrees of freedom of the system. Therefore, instead
of following that approach, we will directly employ
our knowledge of the formal solutions to the
densitized Hamiltonian constraint, deduced in the
previous subsection, and determine a physical Hilbert
structure for them by requiring the self-adjointness
of a complete set of real observables.

Solutions to the constraint are formally given by
expression \eqref{solution} with $\rho=0$. As we can
see, the initial data
$(\psi_{\lambda_\sigma,\lambda_\delta}(\varepsilon_\theta)|$
completely determines the formal solution, so that we
can identify the latter with the data on the initial
section $v_\theta=\varepsilon_\theta$. To construct
the physical Hilbert space, we then simply provide the
vector space of initial data with a Hilbert structure.
The desired inner product can be fixed by choosing a
complete set of real classical observables and
requiring that their quantum analogs, which act on the
initial data, be self-adjoint operators. Such a
complete set is provided by the observables introduced
for Bianchi I in Sec. \eqref{subsec:obs} (with a
trivial action on the inhomogeneous sector) and by a
complete set acting on the inhomogeneous modes, which
can be, for instance, the set of operators that
represent the Fourier sine and cosine coefficients of
the nonzero modes. Up to irrelevant constant real
factors, these operators are
\begin{equation}\label{realF}
\left\{(\hat a_m+\hat a_m^\dagger)\pm(\hat a_{-m}+
\hat a_{-m}^\dagger),\quad i[(\hat a_m-\hat
a_m^\dagger)\pm(\hat a_{-m}- \hat a_{-m}^\dagger)];
\quad m\in\mathbb{N}^+ \right\}.
\end{equation}
Actually, they are self-adjoint operators in the
standard Fock space $\mathcal F$. Therefore, we
conclude that the Hilbert space of initial data picked
up by our conditions is, up to equivalence,
$L^2(\mathbb{R}^2,
|\lambda_\sigma+\lambda_\delta|d\lambda_\sigma
d\lambda_\delta) \otimes\mathcal F$.

Finally, to obtain the true physical Hilbert space of
the model, we still have to impose the $S^1$-symmetry
generated by the constraint $\widehat C_\theta$, which
commutes with the Hamiltonian constraint. This
symmetry is encoded in the condition
\eqref{S1symmetry}. Taking this into account, we
arrive in the end to the physical Hilbert space
\begin{equation}\label{physpaceG}
\mathcal H^{\text{G}}=L^2\left(\mathbb{R}^2,
|\lambda_\sigma +\lambda_\delta| d\lambda_\sigma
d\lambda_\delta\right)\otimes \mathcal F_p.
\end{equation}

\section{Concluding Remarks}
\setcounter{subsection}{0}

We have carried out a thorough quantization of the
Gowdy spacetimes with the spatial section of a
three-torus and linear polarization by combining the
loop quantization of the homogenous Bianchi I
cosmology (with compact sections) \cite{mmp} with the
Fock quantization of the inhomogeneities in a totally
deparametrized Gowdy system \cite{men1}. In this way,
we have constructed a hybrid quantum model for this
family of cosmological spacetimes in vacuo which
incorporates the presumably most relevant effects of
the quantization of the geometry (at least in
scenarios that can be considered close to
homogeneity), while allowing the treatment of an
infinite number of degrees of freedom.

Even though the Hamiltonian constraint couples in a
nontrivial manner the polymeric quantum homogeneous
sector with the inhomogeneous sector, quantized with
standard Fock methods, we have been able to find the
formal solutions to this constraint. The constraint
can be regarded as an evolution equation in an
emergent time, and the explicit expression of the
solutions shows that they are in fact completely
determined by their data at an initial value of this
time.

The emergent time corresponds to one of the variables
that have been quantized with loop techniques. Because
of this fact, a naive and straightforward definition
of the evolution with respect to it is not implemented
as a unitary transformation in the quantum theory. We
note, in this sense, that the results discussed at the
end of Sec. \eqref{subsec:obs} for Bianchi I may be
extrapolated (at least in principle) to the case of
the hybrid Gowdy model, because this model admits
observables which are just the tensor product of
Bianchi I observables and the identity on the
inhomogeneous sector.

Owing to the polymeric quantization carried out in the
homogeneous sector, we have been able to decouple the
space of states in the kernel of any of the triad
operators and rigorously remove it from our
quantization. Therefore, there is no longer a quantum
analog of the (generic) classical cosmological
singularity in the system. This result, originally
obtained for the Bianchi I model, extends also to the
Gowdy model. Hence, we conclude that the loop
quantization of the considered zero modes suffice to
avoid the singularity in this inhomogeneous model.
This robust result is different and simpler than other
suggested possibilities for avoiding cosmological
singularities in inhomogeneous scenarios, like e.g. by
appealing to the BKL (Belinsky, Khalatnikov, and
Lifshitz) conjecture \cite{bkl}. In particular, the
persistence of the resolution mechanism found in
homogenous LQC is a global result, in the sense that
one does not need to analyze the approach to the
singularity independently at each point of the
corresponding spatial section. Furthermore, no
particular boundary condition has to be imposed in the
construction of the quantum states in order to avoid
the singularity and prevent the emergence of
contributions with different triad orientation. From
this perspective, one can say that the developed
formalism provides a no boundary description.

In addition to all the above, we have been able to
complete the hybrid quantization by determining the
Hilbert space of physical states and providing a
complete set of observables acting on it. Remarkably,
the physical Hilbert space which results of this
hybrid quantization is (equivalent to) the tensor
product of the physical Hilbert space for Bianchi I in
LQC and the Fock space which describes the
quantization of the inhomogeneities in the totally
deparametrized system \cite{men1}. Therefore, we can
say that one recovers the standard quantum field
theory for the inhomogeneities on a polymerically
quantized Bianchi I background. Finally, let us
mention that there are other interesting issues for
future investigation for which our approach may be
specially appropriate, like e.g. the analysis of the
semiclassical behavior of physical states, or the
implementation of perturbative approaches which deal
with the inhomogeneities as perturbations around the
Bianchi I background.

\section*{Acknowledgments}

The authors are grateful to J.M. Velhinho, L.J. Garay,
and T. Pawlowski for discussions. This work was
supported by the Spanish MICINN Project
FIS2008-06078-C03-03 and the Consolider-Ingenio 2010
Program CPAN (CSD2007-00042). M. M-B. acknowledges
financial aid by CSIC and the European Social Fund
under the grant I3P-BPD2006.


\begin{thebibliography}{0}    

\bibitem{hawel} S.W. Hawking and G.F.R. Ellis,
{\it{The Large Scale Structure of Space-Time}}
(Cambridge University Press, Cambridge, England,
1973).

\bibitem{lqg} T. Thiemann, {\it{Modern Canonical
Quantum General Relativity}} (Cambridge University
Press, Cambridge, England, 2007); C. Rovelli,
{\it{Quantum Gravity}} (Cambridge University Press,
Cambridge, England, 2004); A. Ashtekar and J.
Lewandowski, Classical Quantum Gravity {\bf 21}, R53
(2004).

\bibitem{lqc} M. Bojowald, Living Rev. Rel. {\bf8}, 11
(2005).

\bibitem{abl} A. Ashtekar, M. Bojowald, and J.
Lewandowski, Adv. Theor. Math. Phys. {\bf7}, 233
(2003).

\bibitem{aps1} A. Ashtekar, T. Pawlowski, and P. Singh,
Phys. Rev. Lett. {\bf 96}, 141301 (2006); Phys. Rev. D
{\bf 73}, 124038 (2006).

\bibitem{aps3} A. Ashtekar, T. Pawlowski, and P. Singh,
Phys. Rev. D {\bf 74}, 084003 (2006).

\bibitem{iso} See e.g., A. Ashtekar, T. Pawlowski,
P. Singh, and K. Vandersloot, Phys. Rev. D {\bf 75},
024035 (2007); L. Szulc, W. Kaminski, and J.
Lewandowski, Classical Quantum Gravity {\bf 24}, 2621
(2007); K. Vandersloot, Phys. Rev. D {\bf 75}, 023523
(2007); E. Bentivegna and T. Pawlowski, Phys. Rev. D
{\bf 77}, 124025 (2008).

\bibitem{chio} D.W. Chiou, Phys. Rev. D {\bf75},
024029 (2007).

\bibitem{mmp} M. Mart\'{\i}n-Benito, G.A. Mena
Marug\'{a}n, and T. Paw\-lowski, Phys. Rev. D {\bf78},
064008 (2008).

\bibitem{szu} L. Szulc, Phys. Rev. D {\bf78}, 064035
(2008).

\bibitem{gowd} R.H. Gowdy, Ann. Phys. {\bf83}, 203
(1974).

\bibitem{qGow} See, e.g., C.W. Misner, Phys. Rev. D
{\bf8}, 3271 (1973); B.K. Berger, Ann. Phys. {\bf83},
458 (1974); Phys. Rev. D {\bf 11}, 2770 (1975); Ann.
Phys. {\bf 156}, 155 (1984); G.A. Mena Marug\'{a}n,
Phys. Rev. D {\bf56}, 908 (1997); M. Pierri, Int. J.
Mod. Phys. Phys. D {\bf11}, 135 (2002).

\bibitem{men1} A. Corichi, J. Cortez, and G.A. Mena
Marug\'{a}n, Phys. Rev. D {\bf73}, 041502 (2006), A.
Corichi, J. Cortez, and G.A. Mena Marug\'{a}n, Phys.
Rev. D {\bf73}, 084020 (2006).

\bibitem{men2} A. Corichi, J. Cortez, G.A. Mena
Marug\'{a}n, and J.M. Velhinho, Classical Quantum
Gravity {\bf23}, 6301 (2006); Phys. Rev. D {\bf76},
124031 (2007); J. Cortez, G.A. Mena Marug\'{a}n, and
J.M. Velhinho, Phys. Rev. D {\bf75}, 084027 (2007).

\bibitem{mon} V. Moncrief, Phys. Rev. D {\bf23},
312 (1981).

\bibitem{ise} J. Isenberg and V. Moncrief, Ann. Phys.
{\bf 199}, 84 (1990).

\bibitem{let} M. Mart\'{\i}n-Benito, L.J. Garay, and G.A.
Mena Marug\'{a}n, Phys. Rev. D {\bf78}, 083516 (2008).

\bibitem{man} G.A. Mena Marug\'{a}n, and M. Montejo, Phys.
Rev. D {\bf58}, 104017 (1998).

\bibitem{new} L.J. Garay, M. Mart\'{\i}n-Benito, G.A.
Mena Marug\'an, and J.M. Velhinho (in preparation).

\bibitem{chi2} D.W. Chiou, Phys. Rev. D {\bf 76},
124037 (2007).

\bibitem{note} Here, we do not use the Einstein summation
convention.

\bibitem{gior} G. Immirzi, Nucl. Phys. B (Proc. Suppl.)
{\bf 57}, 65 (1997); Classical Quantum Gravity {\bf
14}, L177 (1997).

\bibitem{vel} J.M. Velhinho, Classical Quantum Gravity
{\bf24}, 3745 (2007).

\bibitem{spectral} M. Reed and B. Simon, {\it{Methods
of Modern Mathematical Physics. I: Functional
Analysis}} (Academic Press, San Diego, 1980).

\bibitem{gave} D. Marolf, \texttt{arXiv:gr-qc/9508015};
Classical Quantum Gravity {\bf 12}, 1199 (1995); {\bf
12}, 1441 (1995); {\bf 12}, 2469, (1995).

\bibitem{gave2}A. Ashtekar, J. Lewandowski, D. Marolf,
J. Mour\~ao, and T. Thiemann, J. Math. Phys. {\bf 36},
6456 (1995).

\bibitem{mmp2} M. Mart\'{\i}n-Benito, G.A. Mena
Marug\'{a}n, and T. Paw\-lowski (in preparation).

\bibitem{bkl} V.A. Belinsky, I.M. Khalatnikov, and E.M.
Lifshitz, Adv. Phys. {\bf31}, 639 (1982).

\end{thebibliography}
\end{document}